# Determination of interaction between a dust particle pair in complex plasmas

Ke Qiao,* Zhiyue Ding, Jie Kong, Mudi Chen, Lorin S. Matthews and Truell W. Hyde†
Center for Astrophysics, Space Physics and Engineering Research, Baylor University, Waco, Texas 76798-7310, USA

A non-intrusive method to measure particle interaction using only the thermal motion of the particles is applied to a vertically aligned dust particle pair in a complex plasma. The scanning mode spectra (SMS) are obtained by tracking the thermal motion of the grains, with the interaction strength then determined from the frequencies and the eigenvector configuration of the normal modes. The interaction of the bottom particle acting on the top particle is shown to be Yukawa-like with the screening length suppressed against the ion flow. The interaction of the top particle acting on the bottom particle is repulsive in the vertical direction and attractive in the horizontal direction. The vertical interaction from the top to bottom particle is stronger than that from the bottom to top particle, agreeing with an extended ion wake tail as predicted by the inhomogeneous ion wake theory. Determination of the horizontal attraction strength serves as a direct verification and quantification of the ion wake effect. Heating of the lower particle in both the vertical and horizontal directions is observed and quantitatively related to the nonreciprocity of the interaction. The *in situ* confinement strength at the position of the bottom particle is found to be consistently lower than at the position of the top particle, caused by discharging of the lower particle by 10-30% while inside the top particle's ion wake.

Determination of the particle-particle interaction lies at the heart of research on physical systems ranging in size from several atoms to macroscopic bodies. Experimental limitations on both visualization and separation control requires indirect measurement of the interaction between microparticles using methods such as reconstruction from macroscopic forces [1], light or neutron scattering [2], and precision vibrational spectroscopy [3]. For systems comprised of medium sized particles (0.01-$10^3$ μm) such as colloidal suspensions, granular materials and complex (dusty) plasmas, direct measurements are possible. Common investigation methods include atomic force microscopy [4] and optical tweezers [5, 6]. A drawback of each is that these are intrusive methods which alter the particle's natural environment. They are therefore not reliable for measurement of the effective interaction between particles mediated by their ambient medium.

A complex plasma consists of micrometer-sized dust particles immersed in a partially ionized plasma, making it a system in which the medium-mediated interaction plays a particularly important role. The dust particles are negatively charged resulting in strong mutual interaction and interesting dynamical behavior such as crystallization [7-9] and ultra-low frequency waves [10-13]. Due to the dynamic response of the ambient plasma to local electric potentials resulting in density redistribution [9], shadowing effects [14], and ion flow [15-19], the interaction between the particles is far more complex than simple Coulomb repulsion.

One advantage of the medium sized systems is that the particles are large enough to be viewed with video microscopy, allowing their motion to be directly tracked. Experiments in complex plasmas have employed tracking of both the thermal [20, 21] and driven motion [22, 23] to probe the interaction between particles. The former studied the correlation of the thermal motion of a particle pair and provided qualitative results about the ion wake mediated interaction. The latter achieved quantitative measurement of the interaction strength, but involved an external driving voltage that alters the plasma conditions. The measurement in the latter case is also limited to the vertical direction, along the gradient in the applied voltage.

Here a method is presented to determine the particle-particle interaction strength from the thermally excited normal modes in a complex plasma system. The thermal motion of a vertically aligned dust particle pair is tracked and the normal mode spectra are obtained to determine both the resonant frequencies and eigenvector configuration of the modes. Making no assumption about the form of interaction, the interaction strength and *in situ* confinement at each particle position is derived from the mode frequencies and eigenvectors employing a theoretical model of coupled oscillators applicable in both the vertical and horizontal direction.

The experiment was carried out in a Gaseous Electronics Conference rf reference cell [24] with two 8 cm-diameter electrodes separated by a distance of 1.9 cm. The lower electrode is powered at 13.56 MHz while the upper ring-shaped electrode and chamber are grounded. A 20 mm × 18 mm × 18 mm (height × length × width) glass box placed on the lower electrode creates the confinement potential needed to establish the dust pairs. Experiments were conducted in argon plasma at 5.7 Pa employing rf powers of 1.5-10 W. A vertical chain of 8.89-μm melamine formaldehyde particles was formed within the glass box with the lower particles dropped by decreasing the rf power until a two-particle vertical pair was left [25, 26]. A vertically fanned laser sheet illuminated the particles and side view images were recorded for 30 s using a CCD camera at 250 fps. The resulting series of images was analyzed to obtain each particle's position and velocity due to thermal fluctuation.

In standard normal mode analysis, the time series of particle velocities is projected onto the eigenvectors corresponding to the modes, which are derived from a presumed form of the particle interaction [27-30]. In the current research, we desire to *determine* the particle interaction. Therefore we scan the eigenvector space, varying the ratio σ of the oscillation amplitudes of the top and bottom particles from ∞ to -∞ to obtain the spectral power density for each of these eigenvectors, a method we refer to as the scanning mode spectra (SMS). The phase angle $\alpha$ is defined as $\sigma = \cos(\alpha)/\sin(\alpha)$ (where $\alpha$ varies between 0 and π as σ varies between ∞ and -∞) (Fig. 1 (a)). The SMS for a particle pair obtained from the vertical and horizontal velocities both exhibit two spectral lines, corresponding to two normal modes with frequencies $\omega_+$ (higher frequency) and $\omega_-$ (lower frequency) (Fig. 1 (b, c)). The maxima in the two spectral lines correspond to $\sigma_+$ and $\sigma_-$, the ratio between the oscillation amplitudes of the top and bottom particles, identifying the eigenvector configuration corresponding to each mode.

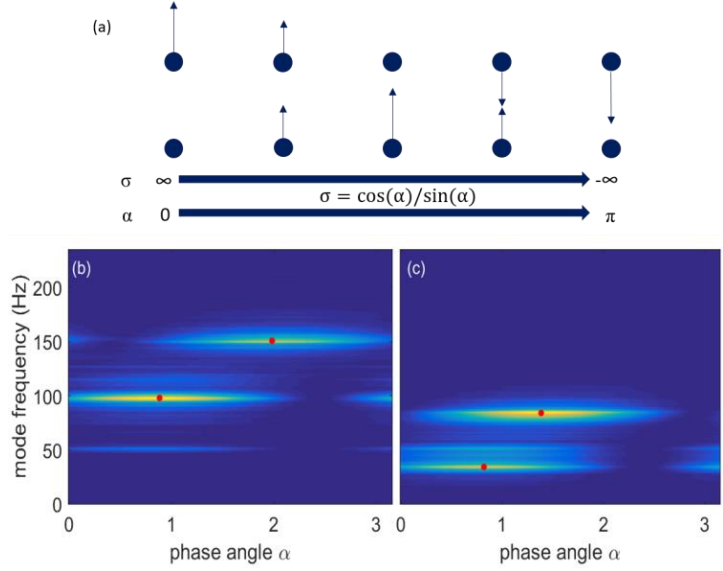

FIG. 1. (Color online) (a) Eigenvector space scanned by varying the ratio σ from ∞ to -∞ (0 ≤ $\alpha \leq \pi$). (b-c) SMS for a particle pair at rf power of 5.5 W obtained from the (b) vertical and (c) horizontal velocity. Red solid dots indicate the maxima in the spectral lines.

Considering the particle pair as two linearly coupled oscillators, the equations of motion are

$$\ddot{x}_1 = -\omega_1^2 x_1 - D_{21}(x_1 - x_2) \quad (1)$$
$$\ddot{x}_2 = -\omega_2^2 x_2 - D_{12}(x_2 - x_1) \quad (2)$$

where $x_{1(2)}$ is the displacement of the top(bottom) particle from equilibrium. The *in situ* confinement at these particle positions determine the frequencies $\omega_{1,2}$, while $D_{12(21)}$ represents the linearized interaction strength of particle 1(2) acting on particle 2(1), defined as the derivative of the interparticle force with respect to displacement [22]. Since the particle motion is due to thermal fluctuations, the displacements are small enough that the forces can be considered linear even when the confinement at the equilibrium positions is not harmonic or the particle charge varies with height. Gas friction is balanced by the random thermal kick on the particles, which together generate the random but saturated thermal fluctuation of the particle system allowing the mode spectra to be obtained. Since these forces do not affect the value of the mode frequencies or eigenvectors [31], they are not included in Eqs. (1) and (2).

Two normal modes can be derived from Eqs. (1) and (2), with frequencies

$$\omega_\pm^2 = \frac{\omega_1^2 + D_{21} + \omega_2^2 + D_{12} \pm \sqrt{(\omega_1^2 - \omega_2^2)^2 + (D_{21} + D_{12})^2 + 2(\omega_1^2 - \omega_2^2)(D_{21} - D_{12})}}{2} \quad (3)$$

and the ratios of the amplitudes σ₊ and σ₋

$$\sigma_\pm = \frac{\left((\omega_1^2 - \omega_2^2) + (D_{21} - D_{12}) \mp \sqrt{(\omega_1^2 - \omega_2^2)^2 + (D_{21} + D_{12})^2 + 2(\omega_1^2 - \omega_2^2)(D_{21} - D_{12})}\right)}{2D_{21}}. \quad (4)$$

Solving Eqs. (3) and (4) for $\omega_1^2$, $\omega_2^2$, $D_{12}$ and $D_{21}$ yield the relationships,

$$D_{21}^2 = \frac{(\omega_+^2 - \omega_-^2)^2}{(\sigma_- - \sigma_+)^2} \tag{5}$$

$$\frac{D_{12}}{D_{21}} = -\sigma_+ \sigma_-. \tag{6}$$

$$\omega_1^2 = \frac{D_{21}(\sigma_+ + \sigma_-) + (\omega_+^2 + \omega_-^2)}{2} - D_{21} \tag{7}$$

$$\omega_2^2 = \frac{(\omega_+^2 + \omega_-^2) - D_{21}(\sigma_+ + \sigma_-)}{2} - D_{12}. \tag{8}$$

As shown, the above allows the interaction strengths $D_{12}$, $D_{21}$ and *in situ* confinements $\omega_1$ and $\omega_2$ to be derived from the experimentally determined values of $\omega_+$, $\omega_-$, $\sigma_+$ and $\sigma_-$. Notice that this model can be applied to motion in either the vertical or horizontal direction, allowing both to be treated in a consistent manner. (Parameters corresponding to the vertical (horizontal) direction are denoted by a subscript $z$ ($x$).)

The interaction strength due to a repulsive Coulomb potential is

$$D_{12z} = D_{21z} = \frac{q_1 q_2}{2\pi\varepsilon_0 m r^3} \tag{9}$$

and

$$D_{12x} = D_{21x} = -\frac{q_1 q_2}{4\pi\varepsilon_0 m r^3} \tag{10}$$

with $q_{1,2}$ the particle charge, $m$ the particle mass and $r$ the interparticle distance. According to Eq. (4), the system will exhibit sloshing ($\sigma > 0$, $\alpha < \pi/2$) and breathing ($\sigma < 0$, $\alpha > \pi/2$) modes in both the vertical and horizontal directions. If the *in situ* confinement $\omega_{1z} = \omega_{2z}$ and $\omega_{1x} = \omega_{2x}$, then $\sigma_{+z} = -1$, $\sigma_{-z} = 1$, $\sigma_{+x} = 1$ and $\sigma_{-x} = -1$. In other words, the breathing mode has a higher frequency than the sloshing mode in the vertical direction, with the relationship reversed for the horizontal direction. The oscillation amplitudes of both particles are equal in the two directions.

The vertical mode spectrum shown in Fig. 1 (b) differs from this ideal case. It exhibits spectral lines at frequencies $\omega_{+z} = 151$ s$^{-1}$ and $\omega_{-z} = 98$ s$^{-1}$, with ratios $\sigma_{+z} = -2.13$ ($\alpha_{+z} = 2.01$) and $\sigma_{-z} = 1.21$ ($\alpha_{-z} = 0.88$). This is indicative of a breathing mode with a higher frequency than the sloshing mode, similar to the ideal case, but $\sigma_{\pm z}$ deviates from $\mp 1$. The derived values, $D_{12z} = 10124$ s$^{-2}$ and $D_{21z} = 3941$ s$^{-2}$ are both positive with a ratio $D_{12z}/D_{21z} \approx 2.57$, showing that the interaction between the two particles is a mutually repulsive Coulomb-like potential (Eq. (9)), but nonreciprocal [22]. From Eq. (4) it can be seen that deviation of $\sigma_{\pm z}$ from $\mp 1$ can be caused by either the nonreciprocity, $D_{12z} \neq D_{21z}$, or a difference in confining strength, $\omega_{1z} \neq \omega_{2z}$. Equal confining strengths imply ratios of $\sigma_{-z} = 1$ and $\sigma_{+z} = -D_{12z}/D_{21z}$. Therefore, the observed deviation of $\sigma_{-z}$ from 1 is the result of the fact that $\omega_{2z}$ is less than $\omega_{1z}$ ($\omega_{1z}^2 = 10395$ s$^{-2}$, $\omega_{2z}^2 = 7823$ s$^{-2}$). Since the value of $\sigma_{+z}$ is nearly equal to $-D_{12z}/D_{21z}$, the difference in the oscillation magnitudes is mainly due to the nonreciprocity of the interaction.

The horizontal mode spectrum (see Fig. 1 (c)) exhibits two sloshing modes [21] with frequencies $\omega_{+x} = 83$ s$^{-1}$, $\omega_{-x} = 34$ s$^{-1}$ and ratios $\sigma_{+x} = 4.47$ ($\alpha_{+x} = 1.35$) and $\sigma_{-x} = 1.06$ ($\alpha_{-x} = 0.82$). This result is completely different from expected values for the Coulomb case. From Eq. (6), the existence of two sloshing modes is the result of $D_{12x} = 7984$ s$^{-2}$ and $D_{21x} = -1676$ s$^{-2}$ having opposite signs. The negative $D_{21x}$ shows that the interaction from the bottom to top particle is a Coulomb-like repulsive force (Eq. (10)). The positive $D_{12x}$, however, shows that the interaction from the top to bottom particle is attractive. As can be seen from Eq. (2), a positive value of $D_{12x}$ indicates that the confinement of the bottom particle is within a horizontal potential well positioned directly below the top particle; therefore, this serves as direct verification and quantification of the ion wake effect. Similar to the results found for the vertical case, the slight deviation of $\sigma_{-x}$ from unity is the result of $\omega_{1x} > \omega_{2x}$ ($\omega_{1x}^2 = 1039$ s$^{-2}$, $\omega_{2x}^2 = 661$ s$^{-2}$) while the large difference in the oscillation magnitudes, $\sigma_{+x} = 4.47$, is caused by the nonreciprocity of the horizontal interaction, $D_{12x}/D_{21x} \approx -4.76$. These results provide a quantitative measurement of the mechanism which has been proposed to cause the heating observed for the lower particle in both the vertical and horizontal directions [21]. The nonreciprocity of the particle interaction may be the origin of more complicated phenomena such as ion flow induced instabilities in a bilayer system [32, 33].

This experimental procedure was repeated, decreasing the rf power from 10 to 1.5 W while keeping the pressure constant at 5.7 Pa. During the process, the separation between the particle pair increases while the center of mass moves first upward and then down for rf power < 2 W. At a power of 1.5 W, the particle separation reaches a critical point, defined by the fact that any further decrease in rf power causes the bottom particle to drop.

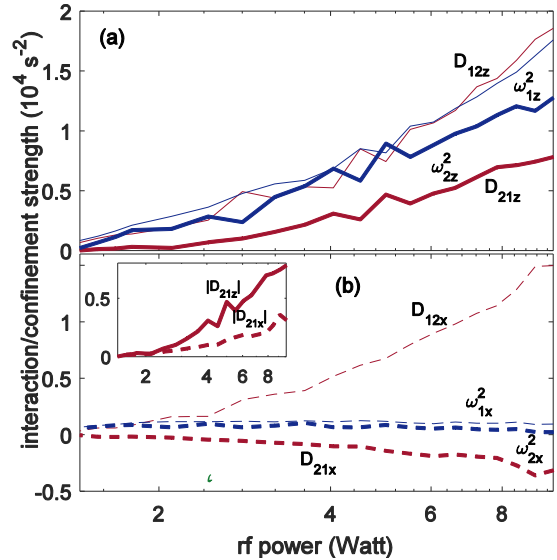

FIG. 2. (Color online) Interaction strengths $D_{12}$, $D_{21}$ and *in situ* confinements $\omega_1$ and $\omega_2$ in the (a) vertical and (b) horizontal directions. Inset: comparison of magnitude of $D_{21z}$ and $D_{21x}$.

As shown in Fig. 2, the magnitudes of $D_{12z,x}$, and $D_{21z,x}$ decrease as power decreases, indicating that the interparticle interaction becomes weaker in both directions as the two particles move farther apart. The repulsive/attractive *nature* of the interaction, indicated by the signs, remains unchanged. In particular, the ion wake attraction coefficient $D_{12x}$ varies from about 15 times the external horizontal confinement $\omega_{2x}$ for high powers to almost zero for low powers. This explains why the vertical pair structure remains stable even when the external vertical confinement $\omega_{2z}$ is much stronger than the horizontal confinement $\omega_{2x}$ at high powers. As shown in Fig. 3, the ratios $\sigma_{+z}$, $\sigma_{-z}$ and $|D_{12z}/D_{21z}|$ remain almost constant for rf power > 4W and increase rapidly in magnitude as the power decreases below 4W as $D_{21z}$ decreases to almost zero (< 300s$^{-2}$) (Fig. 2 (a)). However, the ratio $|D_{12x}/D_{21x}|$ fluctuates around a value of 5 across the entire range of power settings.

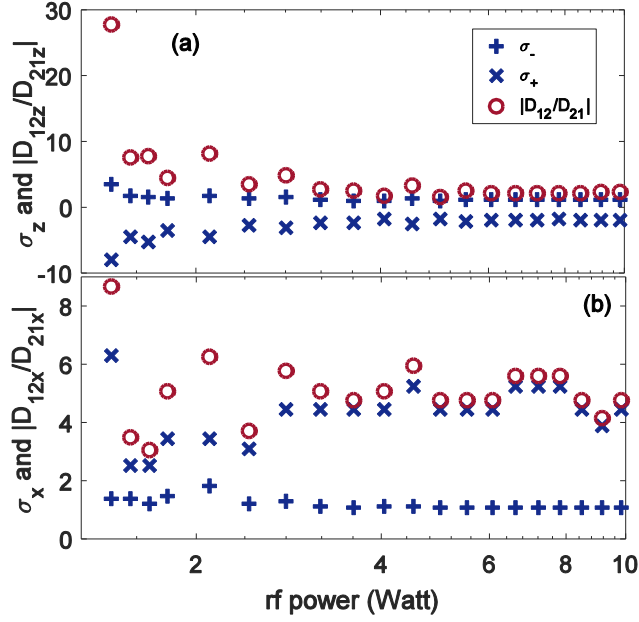

FIG. 3. (Color online) The ratios $\sigma_+$ and $\sigma_-$, corresponding to the two normal modes, and the ratio of nonrecipocity $|D_{12}/D_{21}|$ in the (a) vertical and (b) horizontal direction.

These results can be explained qualitatively by the theoretical electric potential formed around dust particles in a downward ion flow (Fig. 4) [15-19]. The upstream potential is Yukawa-like with the screening length $\lambda$ suppressed along the flow direction, resulting in $\lambda_v$ in the vertical direction being less than $\lambda_h$ in the horizontal direction. The downstream lines of constant potential form nested cones, leading to an attractive horizontal force and a repulsive vertical force. If the ambient plasma is assumed homogeneous [15-17], the nested cones converge, producing a maximum in the potential corresponding to a focused positive space charge region. Considering the sheath inhomogeneity, the cones become divergent, corresponding to a positive space charge region which extends into a long tail (Fig. 4 (a)) [18, 19]. The fact that the particle interaction $D_{12z}$ remains repulsive and greater than $D_{21z}$ over the entire range of particle separation indicates that the wake potential either forms divergent cones, or that the region of convergence is located below the bottom particle. In either case, the downstream potential corresponds to a positive space charge region spread over a long tail, as predicted in inhomogeneous ion wake theory.

At higher powers ($>4$ W), the particles are close enough to one another that their separation distance is less than the upstream screening length $\lambda_v$ of the bottom particle (Fig. 4 (b)), hence the force from bottom to top is Coulombic in nature. As the power decreases below 4 W, the interparticle distance becomes greater than $\lambda_v$ (Fig. 4 (c)), resulting in the decrease of $D_{21z}$ to almost zero (Fig. 2 (a)). This in turn leads to the dramatic increase of nonrecipocity $|D_{12z}/D_{21z}|$ and ratio $\sigma_{+z}$ (Fig. 3 (a)).

Further support for this explanation is provided through estimation of the particle charge and investigation of $D_{21x}$. 1) Using the equilibrium positions of the particle pair and comparing them to that found for a single particle under identical experimental conditions, the charge on the top particle can be calculated from its displacement from equilibrium, assuming a Coulomb repulsion from the bottom particle. The resulting value, $q_1 \approx 12000e$ for rf power $> 4$ W, is in excellent agreement with an independent measurement using the cluster spectra method [29], hence verifying the bottom to top force as Coulombic in nature. 2) As shown in the inset of Fig. 2 (b), for rf powers greater than 4W, $|D_{21x}| \approx |D_{21z}|/2$, as expected for Coulomb repulsion (Eq. (9) and (10)). However, for powers less than 4 W, $|D_{21x}|$ decreases more slowly than $|D_{21z}|$, resulting in $|D_{21x}| \approx |D_{21z}|$. This supports the picture of a screened Coulomb potential with a screening length suppressed in the vertical direction ($\lambda_v < \lambda_h$).

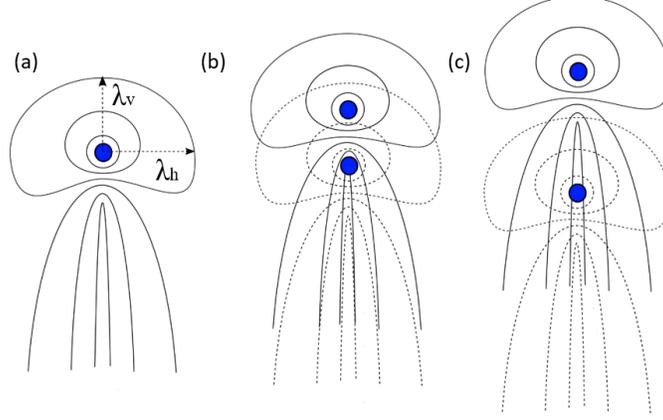

FIG. 4. (Color online) Sketch of the typical theoretical electric potential field formed around (a) one dust particle with ion flow and (b-c) a two-particle pair when their separation is (b) less than and (c) greater than the upstream screening length $\lambda_v$ of the bottom particle.

The relationship between the vertical and horizontal confinement strengths, $\omega_{2z} < \omega_{1z}$ and $\omega_{2x} < \omega_{1x}$, holds over the entire range of rf powers examined. Since $m\omega_{1(2)z}^2$ is defined by the gradient of the electric field force $dF_E/dz|_{1(2)}$, it can be compared to $dF_E/dz$ measured directly by tracking a single dropped particle [34, 35]. The gradient $dF_E/dz$ found by this method is approximately equal at the positions of the two particles. Therefore, the difference between $\omega_{2z}$ and $\omega_{1z}$ must be due to interaction between the particles. One reasonable explanation is the decharging of the lower particle inside the top particle's wake [36]. Since the spatial scale of the field change is typically an order of magnitude less than that of the change in charge [37, 38], the ratio between charges $q_2/q_1$ is approximately equal to $\omega_{2z}^2/\omega_{1z}^2$, which fluctuates between 0.7-0.9. This leads to a decharging ratio of 10-30%, in agreement with previous research [22, 23]. However, $\omega_{2x}/\omega_{1x}$ changes from ~1 to ~0.4 as the power increases, which may be due to an increasing difference in $dE_x/dx$ for the two particles' positions or a change in charge with position $dq/dx$. This is a topic worth future investigation.

In summary, a nonintrusive method employing a scanning mode spectra (SMS) has been used to determine the interaction strength between a particle pair in a complex plasma under constant pressure but varying rf power. The interaction of the bottom particle acting on the top particle is found to be Yukawa-like with a vertical screening length suppressed against the ion flow ($\lambda_v < \lambda_h$). The interaction of the top particle acting on the bottom particle is repulsive in the vertical direction and attractive in the horizontal direction. The vertical interaction strength $D_{12z}$ is larger than $D_{21z}$ for all particle separation distances examined, agreeing with an extended ion wake tail as predicted in inhomogeneous ion wake theory. The horizontal attraction strength $D_{12x}$ serves as direct verification and quantification of the ion wake effect, ranging from 15 times the external confinement $\omega_{2x}^2$ at maximum power to almost zero at minimum power. Heating of the lower particle in both the vertical and horizontal directions was observed and quantitatively related to the nonreciprocity of the interaction. Finally, the *in situ* confinement at the position of the bottom particle is found to be consistently lower than at the position of the top particle. This is caused by decharging of the lower particle inside the top particle's ion wake by 10-30%.

This method is not limited to complex plasmas, but can be applied to any system for which the thermal motion of the particles can be tracked, such as colloidal suspensions and granular materials. Extension of the method to more complex structures with additional constituent particles should also be possible upon application of advanced spectra analysis dealing with multiple dimensions.

This work was supported in part by the National Science Foundation (NSF) under Grant No. PHY 1414523 and PHY 1262031.

*ke_qiao@baylor.edu
†truell_hyde@baylor.edu